\documentclass[fleqn,usenatbib]{mnras}

\usepackage{newtxtext,newtxmath}
\usepackage{soul}
\usepackage{ulem}
\usepackage[T1]{fontenc}

\DeclareRobustCommand{\VAN}[3]{#2}
\let\VANthebibliography\thebibliography
\def\thebibliography{\DeclareRobustCommand{\VAN}[3]{##3}\VANthebibliography}

\usepackage{graphicx}	
\usepackage{amsmath}	
\usepackage{booktabs}
\usepackage{orcidlink}  

\def\apss{Ap\& SS}
\def\apj{ApJ}
\def\apjl{ApJL}
\def\aap{A\& A}

\def\araa{ARA\&A}
\def\mnras{MNRAS}

\def\apjs{ApJS}

\def\mnras{{MNRAS}}

\newcommand{\alphavir}{\alpha_{\rm vir}}

\newcommand{\Sigmagas}{\Sigma_{\rm gas}}

\title[Differential Virial Analysis]{Collapsing clouds need not look virial: a cautionary note on differential virial analysis}

\author[Klos et al.]{
Katerina Klos$^{1}$\thanks{E-mail: k.klos@irya.unam.mx}\orcidlink{}, Javier Ballesteros-Paredes$^{1}$\orcidlink{0000-0002-1081-9445}, Manuel Zamora-Avilés$^2$\orcidlink{0000-0002-2133-9973}, Fabián Quesada-Zúñiga$^2$\orcidlink{0009-0000-3167-2713} 
\\
\\
$^{1}$Universidad Nacional Autónoma de México, Campus Morelia. Instituto de Radioastronom\'ia y Astrof\'isica. PO Box 3-72. 58090. Morelia, Michoac\'an, M\'exico
\\
$^2$ Instituto Nacional de Astrof\'isica, \'Optica y Electr\'onica, Luis E. Erro 1, 72840 Tonantzintla, Puebla, M\'exico
}

\date{Accepted XXX. Received YYY; in original form ZZZ}

\pubyear{\the\year{}}

\begin{document}
\label{firstpage}
\pagerange{\pageref{firstpage}--\pageref{lastpage}}
\maketitle

\begin{abstract}
The dynamical state of molecular clouds is much debated. A key question is whether star-formation proceeds as a result of localised collapse within a globally stable cloud, or due to different centers of collapse within a globally unstable cloud. The virial ratio, $\alpha_{\mathrm{vir}}$, is often used to quantify their stability. However, in practice, interpretation of linewidths, coupled with observational uncertainties makes the {usage} of $\alpha_{\mathrm{vir}}$ problematic. A new technique, known as Differential Virial Analysis (DVA), aims to combat this by tracing the differential change in $\alpha_{\mathrm{vir}}$ as a function of surface density, $\Sigma$. Proponents of this analysis argue that a stable cloud exhibits an $\alphavir$ increasing with $\Sigmagas$ mimicking a hook shape, while unstable clouds exhibit flat $\alphavir-\Sigmagas$ profiles. This method is supposed to distinguish qualitatively between stable versus collapsing systems. We test this method on three realisations of an initially subvirial, collapsing 1000 $M_\odot$ cloud, which evolve just over one free-fall time. In spite of this, we see that the early stages of the cloud's evolution mimic the characteristic ``hook'' shape for $\alpha_{\mathrm{vir}} - \Sigma$ attributed to global stability. As the collapse proceeds, their $\alpha_{\mathrm{vir}}-\Sigma$ profiles transition to approximately flat. Since the evolution of our cloud cannot be attributed to the transition from a hydrostatic to a collapsing system, we argue DVA cannot be taken as a tool to distinguish the inferred dynamical state of molecular clouds. Instead, we suggest it can be used meaningfully in the quantification of the evolutionary state of molecular clouds.
\end{abstract}





\section{Introduction}

{
The dynamical state of molecular clouds remains one of the central questions in the theory of star formation. Molecular clouds exhibit supersonic linewidths over a wide range of spatial scales, and these have traditionally been interpreted as evidence for turbulent non-thermal motions that contribute to supporting the gas against self-gravity \citep[e.g.][]{Larson81, Solomon+87, McKee_Ostriker07, MacLow_Klessen04, Krumholz+19}. In this picture, molecular clouds may be approximately virialized on large scales, while star formation is restricted to the subset of dense regions where turbulent, magnetic, or external-pressure support becomes insufficient.

An alternative interpretation is that the observed non-thermal motions do not primarily represent support, but are instead a consequence of gravity-driven evolution. In this view, molecular clouds are assembled by large-scale flows, become gravitationally unstable as they accumulate mass, and subsequently collapse in a hierarchical fashion, with a number of different smaller and denser structures collapsing inside \citep[e.g.][]{Vazquez-Semadeni+07, Heitsch_Hartmann08, Heitsch+09, Ballesteros-Paredes+11a, Hartmann+12, Camacho+16, Ibanez-Mejia+16, Ibañez-Mejía+22, Vazquez-Semadeni+19}. The observed linewidths may then contain contributions from turbulence, accretion, infall, and inertial motions. Thus, the observational distinction between turbulent support and gravitational contraction is not simply a question of measuring the amount of kinetic energy, but of identifying the physical origin of the motions that produce the linewidth \citep{Ballesteros-Paredes06}.

The virial parameter is one of the most widely used diagnostics in this context, since it is assumed to measure the virial ratio between the kinetic ($E_k$) and gravitational ($E_g$) energy, 
\begin{equation}
  \alpha_{\rm vir} =
  \frac{2E_k}{E_g}.
  \label{eq:alfavir}
\end{equation}
Values of $\alpha_{\rm vir}$ of order unity are often associated with approximate equilibrium. However, this interpretation is not unique. A time-dependent collapsing system can also develop comparable kinetic and gravitational energies, because gravitational contraction itself feeds the kinetic motions. In such a case, virial-like energy balance is not evidence for hydrostatic equilibrium, but a natural consequence of collapse \citep{Vazquez-Semadeni+07, Ballesteros-Paredes+11a, Ballesteros-Paredes+18, Noriega-Mendoza_Aguilar18}. Moreover, projection effects, cloud geometry, magnetic stresses, surface terms, and the distinction between random and coherent motions all complicate the interpretation of the absolute value of $\alpha_{\rm vir}$ \citep{Ballesteros-Paredes06}.
For a projected structure of effective radius $R_{\rm eff}$, mean surface density $\Sigma$, and one-dimensional velocity dispersion $\sigma$, the observational virial parameter is usually written as
\begin{equation}
  \alpha_{\rm vir,obs} =
  \frac{5 \sigma^2}{\pi G \Sigma R_{\rm eff}} .
  \label{eq:avir_obs}
\end{equation}
Equivalently, this expression may be written in the form
\begin{equation}
  \frac{\sigma^2}{R_{\rm eff}} =
  \frac{\pi G}{5}\,\alpha_{\rm vir,obs}\,\Sigma .
  \label{eq:virial_diagram}
\end{equation}
Thus, in the so-called \citet{Heyer+09} diagram, defined by $\sigma^2/R_{\rm eff}$ versus $\Sigma$, lines of constant $\alpha_{\rm vir,obs}$ have slope unity. The location of a structure relative to these lines is commonly used as an estimate of the relative importance of kinetic motions and self-gravity.

The absolute value of $\alpha_{\rm vir,obs}$ is, however, difficult to interpret dynamically. Equation~(\ref{eq:avir_obs}) is an estimator, not a direct measurement of the full virial balance. It assumes a particular geometrical normalization, uses projected quantities, and does not include magnetic terms, surface stresses, external gravity, or the distinction between coherent and random motions, all of which may be important \citep[e.g., ][]{Ballesteros-Paredes+99a, Ballesteros-Paredes06, Ganguly+24}. And as we have already mentioned, the origin of, the line-of-sight (LOS) velocity dispersion entering the numerator may contain contributions from numerous sources. For this reason, a value $\alpha_{\rm vir,obs}\sim1$ does not by itself imply hydrostatic equilibrium, and a value $\alpha_{\rm vir,obs}>1$ does not by itself demonstrate that the gas is supported against collapse.

Recently, \citet{Krumholz+25} proposed a new method, Differential Virial Analysis (DVA), to reduce these ambiguities by considering not only the value of $\alpha_{\rm vir,obs}$ for a whole cloud, but its variation within the same object. Operationally, one selects a set of structures at a defined column density threshold within a global system and computes, for each level, the corresponding values of $\Sigma$, $R_{\rm eff}$, $\sigma$, and $\alpha_{\rm vir,obs}$. The resulting locus in the virial diagram then describes how the apparent ratio between kinetic motions and self-gravity changes as progressively denser gas is selected.

In their interpretation, if a cloud is globally collapsing, then the kinetic motions on different scales are expected to be fed primarily by the gravitational potential energy released during collapse. In that case, structures selected at different surface densities should remain close to lines of constant $\alpha_{\rm vir,obs}$ in the virial diagram, or equivalently $\alpha_{\rm vir,obs}$ should vary only weakly with $\Sigma$. By contrast, if a cloud is globally supported and locally stable, then progressively denser substructures are expected to be less self-gravitating over most of the cloud's dynamic range. In this case, $\alpha_{\rm vir,obs}$ should increase toward higher $\Sigma$, and the trajectory in the virial diagram should bend upward relative to lines of constant $\alpha_{\rm vir,obs}$. 

This argument can be stated as a proposed mapping between the differential virial trend and the underlying dynamical state, where 
\begin{equation}
  \frac{d\alpha_{\rm vir,obs}}{d\Sigma} \simeq 0
\end{equation}
implies some sort of collapse, and
\begin{equation}
  \frac{d\alpha_{\rm vir,obs}}{d\Sigma} > 0
\end{equation}
implies global support.

The differential approach is appealing because it focuses on relative trends within the same object, and is therefore, less sensitive to some of the systematic uncertainties that affect conventional virial analysis \citep{Krumholz+25}. However, its use as a diagnostic of dynamical state requires an additional assumption: that the trajectory followed by a cloud in the virial diagram is unique to its underlying physical state. In other words, an increasing $\alpha_{\rm vir}$ toward high $\Sigma$ must be produced only by supported clouds, while globally collapsing clouds must be unable to produce such a trend.

This uniqueness is, however, not guaranteed. The observed virial parameter combines a projected column density, an effective radius, and a LOS velocity dispersion into a single scalar quantity.

Furthermore, (differential or not) virial diagnostics implicitly assume that observed linewidths trace turbulent, quasi-stationary internal support rather than transient or converging flows. Yet molecular clouds are highly dynamic systems in which  turbulent energy dissipates in a crossing time unless continuously replenished \citep{MacLow+98, Stone+98, MacLow99}, making long-lived turbulent hydrostatic configurations difficult to sustain. In addition, molecular clouds that actively form stars must necessarily host locally collapsing structures. If increasing virial parameter toward high column density is interpreted as evidence against local collapse, then dense star-forming cores may become difficult to reconcile with the diagnostic itself. 

Moreover, the effective radius depends on the projected morphology of the selected gas, and therefore changes not only with the intrinsic gravitational state of the structure, but also with fragmentation, filamentarity, projection, and the adopted column-density threshold. Finally, in a time-dependent collapsing cloud, dense structures do not all become gravitationally unstable simultaneously. Collapse begins on some scales before others, and the measured differential virial trend may therefore reflect the evolutionary ordering of structures rather than a division between supported and collapsing gas \citep[][see also \citealt{Vazquez-Semadeni+19}]{Naranjo-Romero+15}.

It is therefore useful to ask whether a cloud known to be undergoing gravitational contraction from its full three-dimensional evolution can nevertheless generate a differential virial track similar to that interpreted as evidence for turbulent support. If the answer is yes, then an increasing $\alpha_{\rm vir,obs}$ with $\Sigma$ is not a sufficient condition for excluding global collapse. The differential virial parameter may still be informative, but its interpretation must be supplemented by independent information about the velocity field, the evolutionary state of the cloud, or additional observational diagnostics. In this paper we test the uniqueness of DVA using numerical simulations of three molecular clouds undergoing gravitational contraction from the outset and do not pass through a stage of hydrostatic or virial equilibrium. In \S\ref{Sec:NumericalSimulations} we present the simulations we will analyse. \S\ref{sec:results} presents our results, and in \S\ref{Sec:Discussion} we discuss our results. Finally, conclusions are drawn in \S\ref{Sec:Conclusions}. }

\section{Numerical simulations}\label{Sec:NumericalSimulations}

{For the purposes of this study, we simulate the collapse of three subvirial, purely hydrodynamical, clouds, using the same initial conditions as described in \citet{Klos+25}. The simulations were performed using the Lagrangian Smoothed Particle Hydrodynamics (SPH) code PHANTOM, which solves the discretized forms of the equations of hydrodynamics with self-gravity, given in \citet{Price12}.

\subsection{Gravity}
\label{subsec:grav}
PHANTOM determines the self-gravity between particles by solving Poisson's equation;
\begin{equation}
    \nabla^2 \Phi = 4\pi G\rho(\vec{r})
    \label{eq:poisson}
\end{equation}
to give the acceleration
\begin{equation}
    \vec{a}_{\mathrm{selfgrav}} = -\nabla\vec{\Phi}
    \label{eq:selfgrav}
\end{equation}
PHANTOM splits this expression into two parts, based on `short-' and `long-range' interactions;
\begin{equation}
    \vec{a}_{\mathrm{selfgrav}}^a = \vec{a}_{\mathrm{short}}^a + \vec{a}_{\mathrm{long}}^a
    \label{eq:selfgrav2}
\end{equation}
where the subscript $a$ indicates the value for an individual particle. For short-range interactions, the contribution is calculated by direct summation for nearby particles, whereas the long-range interactions are calculated using a $k$d-tree. For further discussion of this procedure, please refer to \citet{Price+18}, section 2.12.

\subsection{Equation of State}
\label{subsec:EOS}

{\citet{Klos+25} use a modified barotropic equation of state \citep{Larson69}, given by
\begin{equation}
\begin{aligned}
    P= c_{\mathrm{s}}^2 \begin{cases}
        \rho, & \rho < \rho_{\mathrm{a}} \\
        \rho_{\mathrm{a}}\left(\frac{\rho}{\rho_{\mathrm{a}}}\right)^{\gamma_{3}}, & \rho_{\mathrm{a}} < \rho < \rho_{\mathrm{c}} \\
        \rho_{\mathrm{a}}\left(\frac{\rho_{c}}{\rho_{\mathrm{a}}}\right)^{\gamma_{3}}\left(\frac{\rho}{\rho_{\mathrm{c}}}\right)^{\gamma_{1}}, & \rho_{\mathrm{c}} < \rho < \rho_{\mathrm{d}} \\
        \rho_{\mathrm{a}}\left(\frac{\rho_{d}}{\rho_{\mathrm{c}}}\right)^{\gamma_{1}}\left(\frac{\rho_{c}}{\rho_{\mathrm{a}}}\right)^{\gamma_{3}}\left(\frac{\rho}{\rho_{\mathrm{d}}}\right)^{\gamma_{2}}, & \rho > \rho_{\mathrm{d}}
    \end{cases}
\end{aligned}
\end{equation}
where $c_{\rm s}$ is the sound speed, which is constant up to density $\rho_{\mathrm{a}}$. The threshold densities correspond to $\rho_{\mathrm{a}} = 10^{-16}$ g cm$^{-3}$, $\rho_{\mathrm{c}} = 10^{-14}$ g cm$^{-3}$, $\rho_{\mathrm{d}} = 10^{-10}$ g cm$^{-3}$. The adiabatic indices are $\gamma_{1} = 7/5$, $\gamma_{2} = 11/10$, and $\gamma_{3} = 5/3$. See \citet{Lewis+15} and \citet{Klos+25} for further discussion of these values.

The adiabatic regime of the simulation only begins for densities greater than $\rho_{\mathrm{a}}$, which is only a factor of 10 below the sink creation threshold (see \S\ref{subsec:sink}). As such, the simulations are practically isothermal, with a constant sound speed, as less than 2\% of dense gas lies above this threshold value for our most condensed cloud, HDcold, at $t = 0.61$ Myr. 

\subsection{Sink Particles}
\label{subsec:sink}

{Sink particles are formed according to the prescription of \citet{Bate+95}, in order to follow the evolution of the cloud once high-density fragments form. The initial sink creation threshold is $\rho_{\mathrm{crit}}~=~10^{-15}$~g~cm$^{-3}$, and the sink accretion radii correspond to $h_{\mathrm{acc}} = 0.01$ pc. Sink merging is switched on, and they merge conditionally within $3~h_{\mathrm{acc}}$ of each other and unconditionally within $2~h_{\mathrm{acc}}$, the latter of which is the default minimum recommendation when using PHANTOM. Sinks are created once the gas meets a number of physical and numerical criteria which are detailed in \citet{Price+18}.}

\subsection{Initial Conditions}
\label{subsec:initcond}

{Details of the initial conditions can be found in \citet{Klos+25}. Here we only discuss the key parameters of each one of the simulations (see also  Table~\ref{tab:cloud_params}).}

{For the analysis of this contribution, we have simulated the collapse of a 1000~M$_{\odot}$, $R~=~1$~pc cloud, initialised with a homogenous density distribution, and an initial turbulent velocity field. We also compare this cloud with two other clouds of the same mass, radius, and density distribution, described in \citet{Klos+25} (HDV and HDC). The new cloud, which we will refer to as HDvir2, and HDV from \citet{Klos+25} (which we will refer to as HDvir) are identical in every way except for the turbulent seed used to set the initial velocity distribution. Cloud HDC (which we refer to now as HDcold), differs from HDvir only in the strength of the initial turbulence, but with the same turbulent seed, such that their virial parameters are 0.5 for HDcold, and 0.9 for HDvir and HDvir2 (see Table~\ref{tab:cloud_params} for more details).}

The initial turbulent velocity field is calculated using a divergence-free random Gaussian velocity field, such that $P(k) \propto k^{-4}$ \citep{Ostriker+01}. Particle velocities are interpolated from the grid such that $\delta v = \mathcal{M}c_{\mathrm{s}}$, where $\mathcal{M}$ is the mach number. In this case, $\mathcal{M} = 6.59$ for clouds HDvir and HDvir2. In the case of HDcold, $\mathcal{M} = 4.66$. The first two clouds therefore begin slightly subvirial, with $2E_k/E_g = 0.9$, with $E_k$ and $E_g$ been the kinetic and gravitational energies. Cloud HDcold begins with $\alpha_{\mathrm{vir}} = 0.5$. It should be stressed that even if this ratio were unity, such a condition will not imply, by any means, a cloud in equilibrium. To be in equilibrium, detailed force balance needs to be achieved everywhere, and thus, an initial density gradient must be needed to compensate the self-gravity of the cloud.

{The cloud is contained within a uniform low-density box, with a density contrast of 30:1 with the initial cloud, as they formed part of a suite of simulations comparing the evolution of clouds in a magnetised medium (e.g. \citet{Wurster+16, Lewis+17, Wurster+21}). Whilst not necessary to the analysis, this initial condition is unlikely to affect the analysis or interpretation of the DVA technique on the clouds. The box has quasi-periodic boundaries, allowing the SPH particles to interact hydrodynamically, but not gravitationally, across the box. The length of the box sides are $L = 4R_{\mathrm{cl}}$, and thus, in practice, the only SPH particles that cross the boundaries from one side to the other are those of the diffuse medium, not the ones from the cloud.  

\begin{table}
    \centering
    \caption{Initial conditions for the spherical cloud collapse
    simulations. All three runs share the parameters in the upper block and differ
    only in their initial turbulent velocity field: two independent random seeds at
    the same virial parameter ($\alpha_{\mathrm{vir}}=0.9$), and one run with lower
    initial turbulent energy ($\alpha_{\mathrm{vir}}=0.5$).}
    \label{tab:initial_conditions}
    \begin{tabular}{lcc}
        \toprule
        Quantity & Symbol & Value \\
        \midrule
        Cloud mass                          & $M_{\mathrm{cl}}$   & $1000\,\mathrm{M_{\odot}}$ \\
        Number of particles in cloud        & $N_{\mathrm{part}}$ & $8 \times 10^{7}$ \\
        Particle mass resolution            & $m_{\mathrm{p}}$    & $1.25 \times 10^{-5}\,\mathrm{M_{\odot}}$ \\
        Min.\ resolvable fragmentation mass & $2M_{\mathrm{kern}}$& $1.45 \times 10^{-3}\,\mathrm{M_{\odot}}$ \\
        Cloud radius                        & $R_{\mathrm{cl}}$   & $1\,\mathrm{pc}$ \\
        Initial density                     & $\rho_{0}$          & $1.61 \times 10^{-20}\,\mathrm{g\,cm^{-3}}$ \\
        Free-fall time                      & $t_{\mathrm{ff}}$& $0.52\,\mathrm{Myr}$ \\
        Temperature                         & $T$                 & $10\,\mathrm{K}$ \\
        Sound speed                         & $c_{\mathrm{s}}$    & $2.189 \times 10^{4}\,\mathrm{cm\,s^{-1}}$ \\
        Angular velocity ($z$-axis) & $\Omega$ & $10^{-14}\,\mathrm{rad\,s^{-1}}$ \\[1.4em]
        \midrule
        \multicolumn{2}{l}{\textit{Turbulent realisation}} & $\alpha_{\mathrm{vir}}$ \\
        \midrule
        HDvir  & & $0.9$ \\
        HDvir2  & & $0.9$ \\
        HDcold & & $0.5$ \\
        \bottomrule
    \end{tabular}
    \label{tab:cloud_params}
\end{table}

These conditions allow us to compare results with section 3.2 from \citet{Krumholz+25}, in their analysis of collapsing molecular clouds. Our simulations consider one cloud per realisation, each forming $\sim$ 20 cores over the course of one free-fall time. From the initial setup, these clouds, by definition, are \textit{always} collapsing. As such, our clouds should, thus, show a clear collapsing signatures under the framework of DVA. Although these initial conditions are relatively simplistic in their initial setup, we believe they work well as a test-bed for the predictions of DVA, as they do not introduce complicated geometries or effects which would make the interpretation of results more ambiguous.

{ \subsection{Computing $\alphavir$ from numerical simulations}
To calculate the evolution of $\alpha_{\mathrm{vir, obs}}-\Sigma$, we followed the same method as described in section 3.2 of \citet{Krumholz+25}, which we will summarize here for clarity. The four quantities we need are discretised as;
\begin{align}
    &\langle \Sigma \rangle  = \frac{1}{N_{\mathrm{pix}}}\sum_{i\in R} \Sigma_i \label{eq:sigsum}\\
    &\langle \sigma^2 \rangle = \frac{1}{N_{\mathrm{pix}}\langle \Sigma \rangle} \sum_{i\in R} \Sigma_i \sigma_i^2 \label{eq:vdisp2}\\
    &R_{\mathrm{eff}} = \sqrt{\frac{N_{\mathrm{pix}}A_{\mathrm{pix}}}{\pi}} \label{eq:reff}\\
    &\alpha_{\mathrm{vir, obs}} = \frac{5 \langle \sigma^2 \rangle / R_{\mathrm{eff}}}{\pi G \langle \Sigma \rangle}
    \label{eq:avirobs}
\end{align}
 
Here, $N_{\mathrm{pix}}$ is the number of pixels in a region, $R$, $\Sigma_i$ is the column density of pixel $i$, $\sigma_i$ is its velocity dispersion, and $A_{\mathrm{pix}}$ is the area of a pixel.

For this, we also require, for each pixel, the column densities and velocity dispersions, which are defined by
\begin{align}
    &\Sigma = \int_{0}^{L}\rho dx \label{eq:sigint} \\
    &\langle v_{x} \rangle = \frac{1}{\Sigma} \int_{0}^{L} \rho v_{x} dx \label{eq:vxave} \\
    &\sigma^2 = \frac{1}{\Sigma} \int_{0}^{L} \rho \left[c_{\mathrm{s},x}^2 + (v_{x} - \langle v_{x} \rangle)^2 \right] dx
\end{align}
where $c_{\mathrm{s},x}$ and $v_x$ are the sound speed and the velocity along $x$, and $L$ is the length of the box. These equations correspond to equations 30 - 33, and 27 - 29 from \citet{Krumholz+25} respectively, transposed to the code units used by our simulations. Although our simulations are practically isothermal, we have also allowed for a variable $c_{\mathrm{s}}$ along the LOS as some material does cross the adiabatic threshold (see \S\ref{subsec:EOS}), though we have found this makes no discernible difference to the results where we used a constant $c_{\mathrm{s}}$ instead.

We begin by interpolating our SPH data onto a grid, in this case 512$^{3}$ compared to \citet{Krumholz+25} who use a 256$^{3}$ grid, where we make use of the high resolution of our simulations ($\sim 8 \times 10^{7}$ gas particles in each cloud). Our grid spans $-1.25 \leq x, y, z \leq 1.25$~pc, to maximise resolution. This 3D grid is then projected onto a 2D 512$^{2}$ column density map. We then define a series of column density contour levels relative to the mean column density of the simulation, $\langle \Sigma \rangle$, and for each of these levels, we identify connected regions with $\Sigma > \Sigma_{\mathrm{contour}}$, for each projection, $x$, $y$, and $z$. Regions with fewer than $N_{\mathrm{noise}} = 120$ pixels are discarded as noise, where we have scaled $N_{\mathrm{noise}}$ accordingly from 30 in \citet{Krumholz+25}, up by a factor of $(512/256)^{2}$. 

For each of the remaining regions, we then calculate the quantities given in equations (\ref{eq:sigsum}) - (\ref{eq:avirobs}). The regions found for each of the three cardinal directions are then pooled together, and at each contour level, we then take the mass-weighted mean of the regions to plot $\alpha_{\mathrm{vir, obs}}$ and $\sigma^2/R_{\mathrm{eff}}$. We do this for levels with three or more regions as opposed to \citet{Krumholz+25} who do it for levels with five or more regions as we are working within a physically smaller simulation volume, and so have less substructure.
}
{
\section{Results} \label{sec:results}

\begin{figure*}
    \centering
    \includegraphics[width=\textwidth]{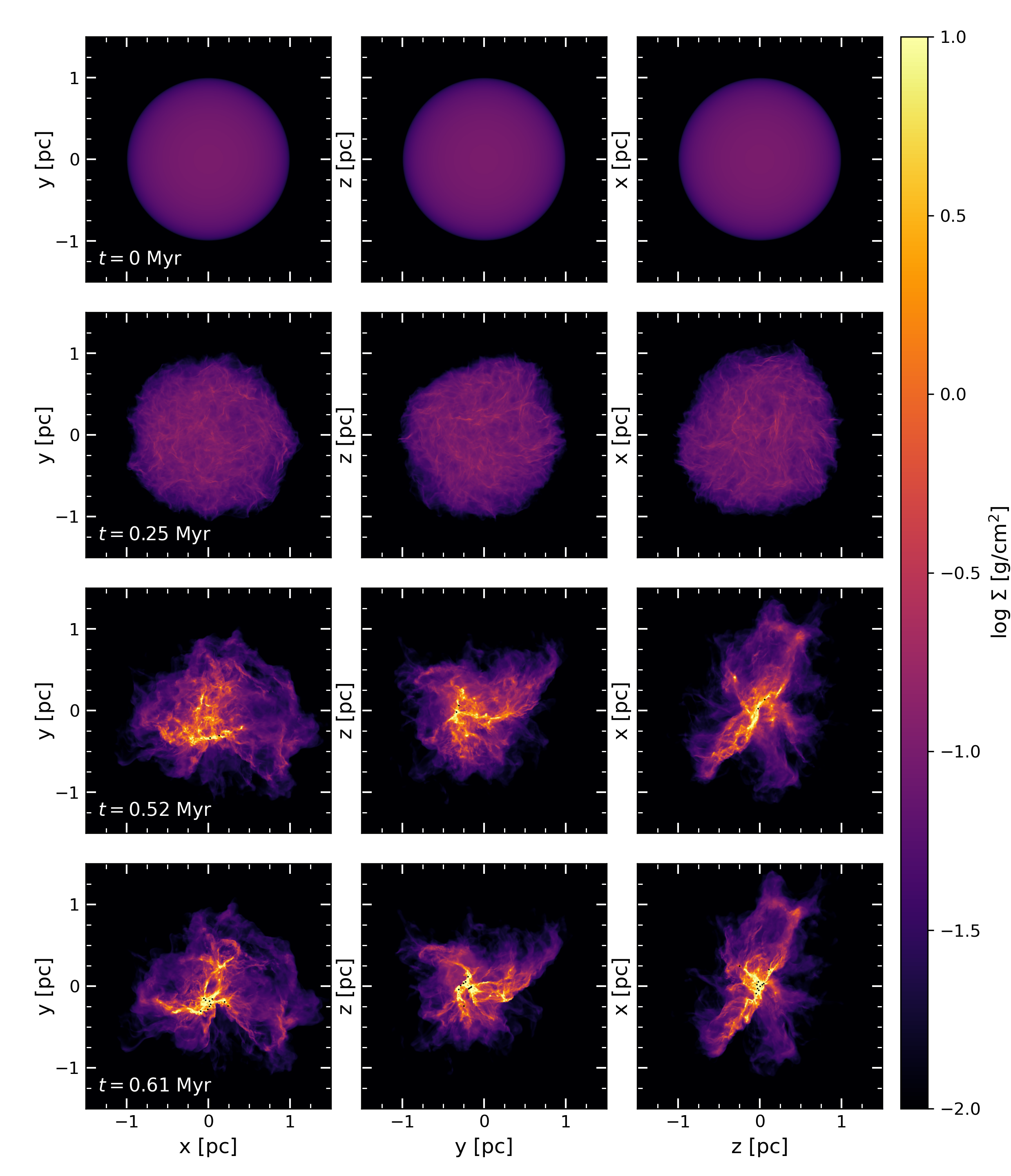}
    \caption{{3D time evolution of cloud HDvir2, as an example, through to $1.2 t_{\mathrm{ff}}$ (0.61 Myr). Each column corresponds to one of three orientations, and each row corresponds to a time. The cloud starts initially spherical, with a uniform density, which shows as an increasing column density towards the centre of the sphere. The second row corresponds to just after one turbulent crossing time, $t_{\mathrm{turb}}$, when the initial turbulent field has time to propagate through the cloud. The third row corresponds to the free-fall time of the initial cloud, $t_{\mathrm{ff}}$, and the last panel is the cloud at the final timestep. By $t_{\mathrm{turb}}$, we see some small-scale structure forming owing to the turbulence, but we also see that the cloud has begun to contract. At $t_{\mathrm{ff}}$, the cloud is significantly deformed in all three dimensions, with clear filamentary structure, and two sinks (black dots). This is further exacerbated in the final row. Whilst much of the internal structure forms due to the turbulent seed, the cloud has unequivocally contracted throughout its full evolution.}}
    \label{fig:3dtimeevol}
\end{figure*}

\begin{figure*}
    \centering
    \includegraphics[width=\linewidth]{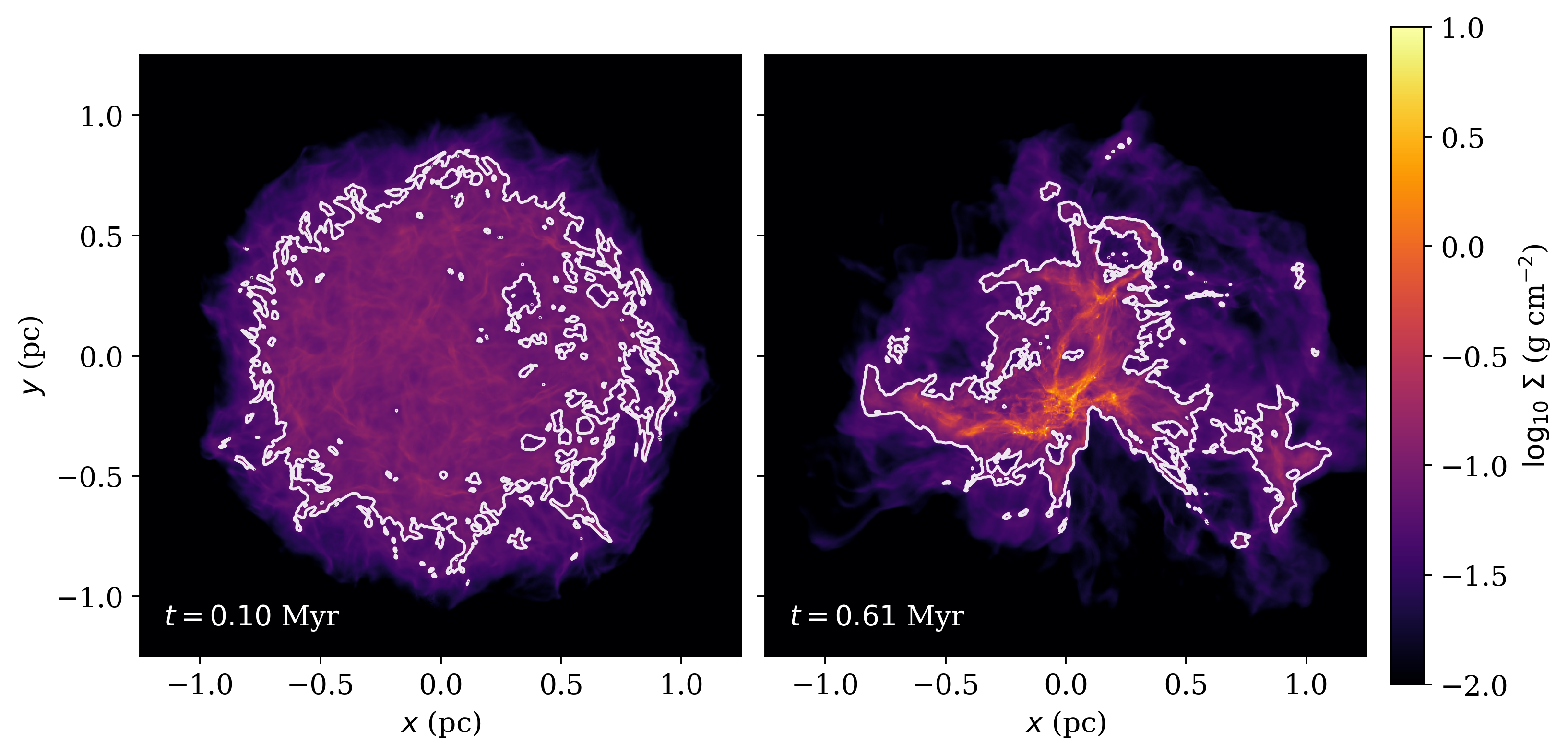}
    \caption{Contour examples for $\log_{10} \Sigma_{\mathrm{thresh}}/\langle \Sigma \rangle$ = 0.25. Here we see the regions along the LOS $z$, identified for DVA analysis in the second timestep, and in the final timestep, for cloud HDvir2. Early on, in the left-hand panel, we see that one dominant structure - tracing the inner region of the cloud - with a few smaller structures identified in the outer regions of the cloud. This reflects the relatively flat global density profile of the cloud, before collapse has really altered the initial conditions of the cloud. In the right-hand panel, we see the cloud in its final, much more centrally condensed state. There is still one large, central structure outlined by the contours, but there are other significant density structures, such as the one centred at $\sim (1.0, -0.5)$, indicating regions of localised collapse.}
    \label{fig:contour_ex}
\end{figure*}

{In Fig.~\ref{fig:3dtimeevol} we show the evolution of the three projections of the density field of cloud HDvir2 (similar images of clouds HDvir and HDcold can be found in \citet{Klos+25}). As can be seen, the cloud starts initially spherical, with a uniform density, which translates to a slightly increased column density towards the centre of the sphere. The second row
corresponds to just after one turbulent crossing time, $t_{\rm turb} = L/\delta v$, with $L$ the size of the cloud and $\delta v$ its velocity dispersion. The third row corresponds
to the free-fall time of the cloud at the initial density, $t_{\mathrm{ff}} = (3\pi/32G\rho)^{1/2}$. The last panel is the cloud at the final timestep, which corresponds to $t=1.2~t_{\mathrm{ff}}$.} 

We now apply the DVA technique described above to our suite of clouds. In figure \ref{fig:contour_ex}, we show an example of the contours drawn for cloud HDvir2 in the first and last timesteps analysed. The contour levels shown are drawn to encompass regions satisfying $\log_{10} \Sigma_{\mathrm{thresh}}/\langle \Sigma \rangle \geq 0.25$. At this level, we see that for the first timestep, the majority of the cloud is encased in one large contour, though there are a few much smaller structures around the cloud edges. This reflects the relative homogeneity of the cloud density distribution earlier on. By the final timestep however, the cloud has shed the initial density distribution, and is much more centrally condensed. The largest contour structure still traces the central regions of the cloud, as would be expected, but there are larger separate components, for example to the bottom right-hand-side of the cloud, indicating more localised regions of collapse.

In Fig.~\ref{fig:VirialTotal} we plot {the 3D-virial parameter (see eq.~\ref{eq:alfavir})} within the clouds, as a function of time, using spherical shell sum for calculating the approximate potential energy of the cloud. This corresponds to
\begin{equation}
    |W| = G\sum_{i}\frac{M_{\mathrm{enc}}(r_{i}) m_{i}}{r_{i}}
    \label{eq:virsum}
\end{equation}
where we have used the sum of spherical shells as an approximation to avoid a pair-wise sum over $\sim 80$ million particles.\footnote{The exact potential energies stored by PHANTOM correspond to the whole box, including the background material.} As such, the approximate virial parameter becomes slightly less reliable at later times as the clouds deviate from spherical symmetry and uniform densities.

As can be seen, clouds HDvir and HDvir2 begin with a slightly subvirial initial condition ($\alphavir\sim 0.9$)  and are thus dominated by gravity. This is even more apparent for cloud HDcold, which has the lowest initial turbulent energy ($\alphavir~=~0.5$).

For the $\alpha_{\mathrm{vir}} \sim 0.9$ clouds, as collapse proceeds, the virial parameter first decreases, as a consequence of the dissipation of the turbulent motions, and then  increases, as a consequence of the collapse itself, reaching values between 1 and 2, a behaviour that can be expected for collapsing systems \citep{Vazquez-Semadeni+07, Ballesteros-Paredes+11a, Ballesteros-Paredes+18, Noriega-Mendoza_Aguilar18}. The collapse of HDcold is more dramatic owing to the lower initial kinetic energy. We see that at later times, owing to the strength of the collapse, the virial parameter climbs up to $\sim 3.3$.

\begin{figure}
 \centering
  \includegraphics[width=\columnwidth]{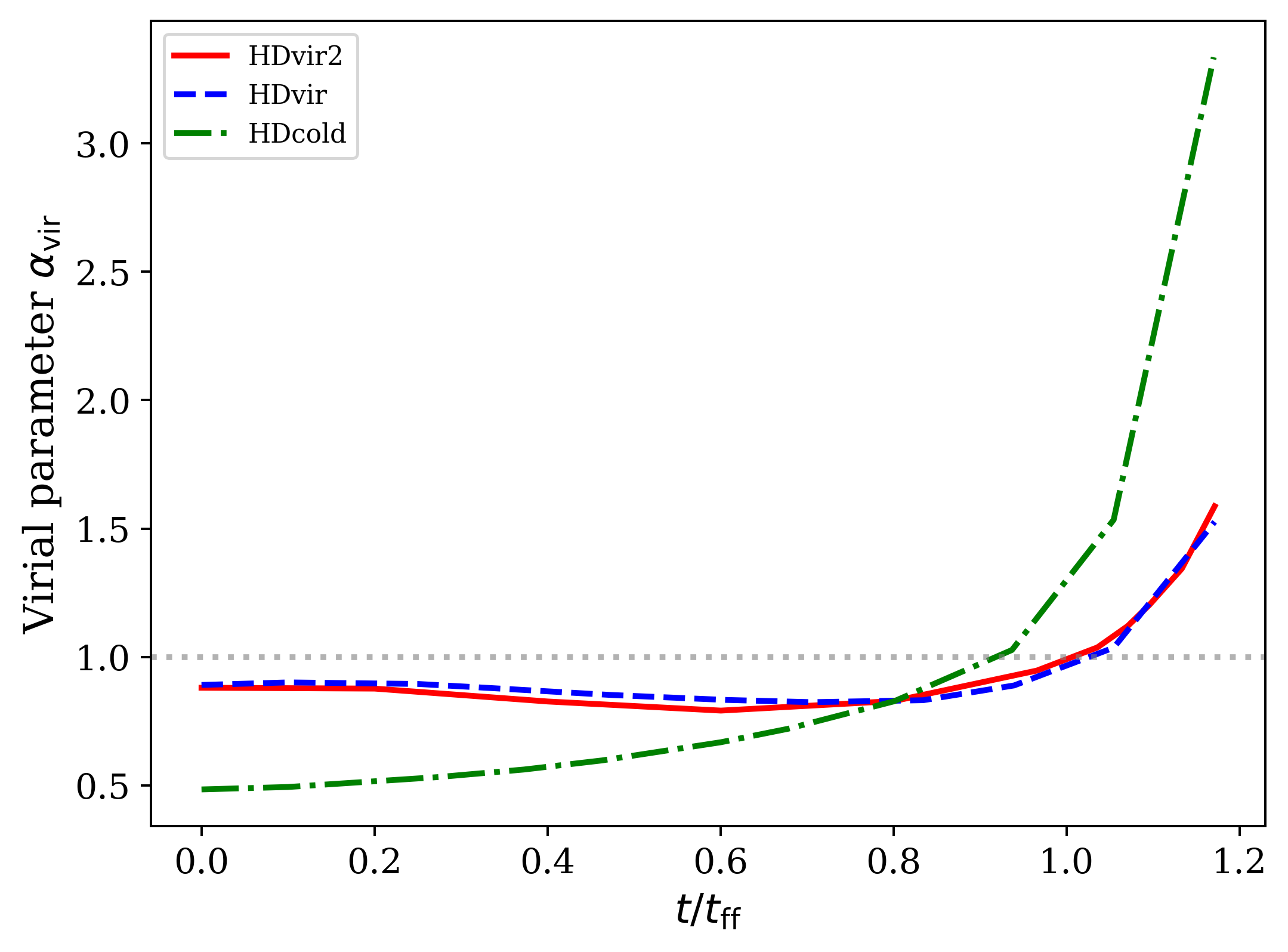}
  \caption{Evolution of the 3D virial parameter for our three clouds as a function of time. The two slightly subvirial clouds begin with initial virial parameters $\sim 0.9$, and show a slight decreased in $\alpha_{\mathrm{vir}}$ before it rises again at later times as the clouds collapse. Cloud HDcold starts with a much lower initial kinetic energy, and collapse proceeds much more freely. As such, $\alpha_{\mathrm{vir}}$ rises much more sharply after about one $t_{\mathrm{ff}}$, ending with $\alpha_{\mathrm{vir}} \sim 3.3$. The kinetic contribution to the virial parameter here is a clear consequence of collapse.}
    \label{fig:VirialTotal}
\end{figure}

One can ask whether the projected quantities used in DVA recover this dynamical state uniquely. Figure~\ref{fig:VirialDiferencialAll}a shows the evolution of the differential virial parameter as a function of projected surface density. Each curve corresponds to a different time, and, following \citet{Krumholz+25}, each point along a curve is obtained by selecting gas above a given surface-density threshold and computing the corresponding values of $\Sigma$, $R_{\rm eff}$, $\sigma$, and $\alpha_{\rm vir,obs}$. Fig.~\ref{fig:VirialDiferencialAll}b shows the corresponding curves in the \citet{Heyer+09} diagram, $\sigma^2/R_{\rm eff}$ versus $\Sigma$. Our Figure~\ref{fig:VirialDiferencialAll} is a recreation of Figure 3 from \citet{Krumholz+25}, for our three simulations. 

\begin{figure*}
    \centering
    \includegraphics[width=\textwidth]{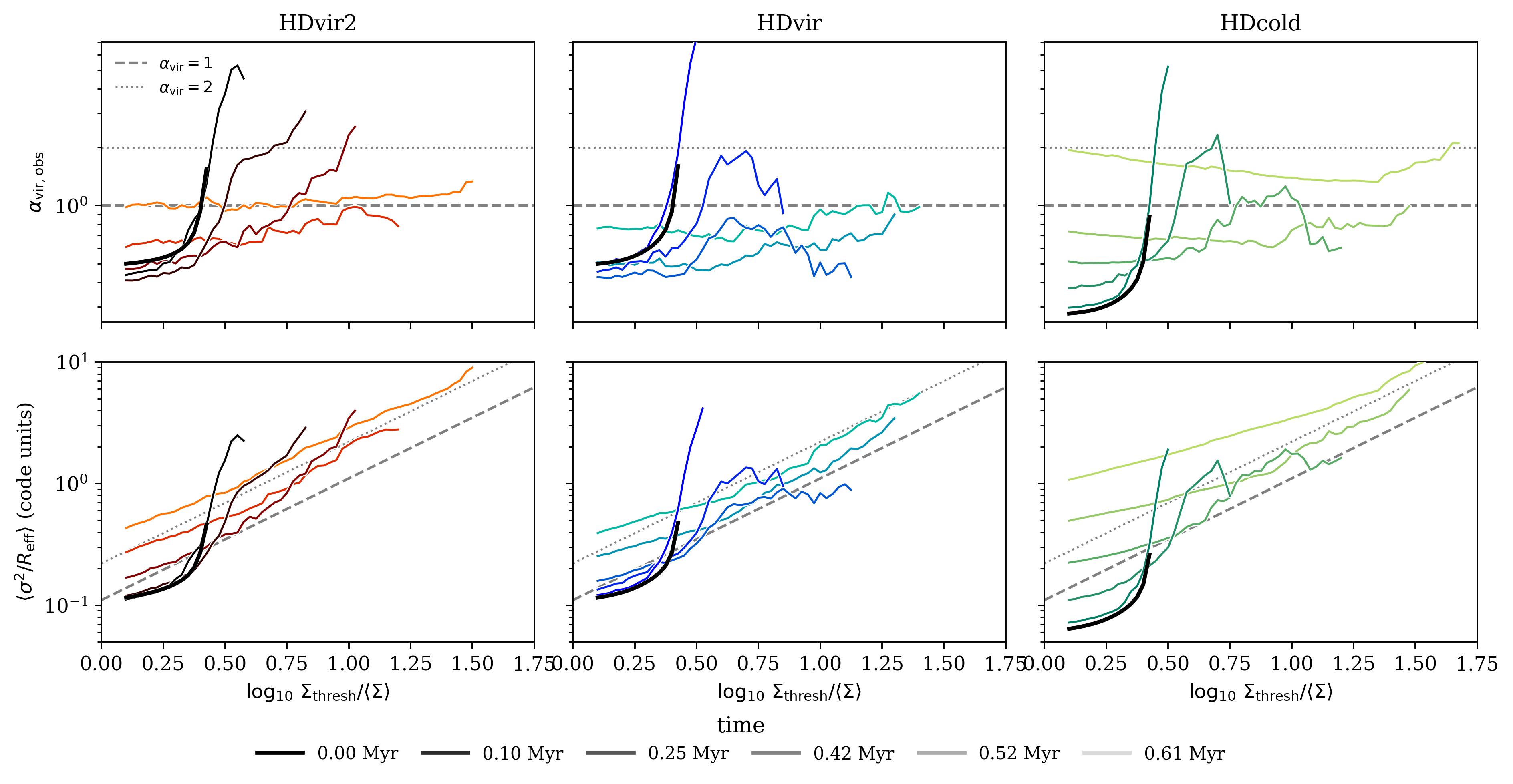}
    \caption{{Differential virial analysis for each of our clouds. The upper panels show the observational virial parameter $\alpha_{\rm vir,obs}$ as a function of projected surface density $\Sigma$ for different evolutionary times. Time is indicated by the shade of the lines, with darker lines indicating earlier times and lighter lines indicating later times. The thick black line in each panel corresponds to $t = 0.0$ Myr. The lower panel shows the same data in the virial diagram, $\sigma^2/R_{\rm eff}$ versus $\Sigma$. Dashed and dotted grey lines indicate $\alpha_{\rm vir,obs}=1$ and $\alpha_{\rm vir,obs}=2$, respectively. Although the clouds are undergoing gravitational collapse, the early tracks for each row all show increasing $\alpha_{\rm vir,obs}$ toward high $\Sigma$, mimicking the behaviour interpreted by DVA as evidence for support. This is also true for the $t = 0$ Myr lines, which follow closely the $t = 0.1$ Myr lines for all the clouds. The progressive flattening of the tracks at later times therefore does not correspond to a transition from support to collapse, but to the evolution of an already collapsing system. Flattening happens at about $\sim t_{\mathrm{ff}}$ of the cloud (which is $\sim 0.52$ Myr). Adding in the effect of the initial turbulence pushes the approximate total collapse timescale to $\sim 0.61$ Myr (the final timestep).}}
    \label{fig:VirialDiferencialAll}
\end{figure*}

As can be seen, although all three clouds are, at all times, undergoing gravitational collapse, it is not until about one free-fall time that they begin to exhibit flat differential virial profiles. Clouds HDvir and HDvir2 show $\alpha_{\mathrm{vir,obs}} \sim 0.5$, whereas the cold cloud has $\alpha_{\mathrm{vir,obs}} \sim 0.3$, reflecting the lower initial kinetic energy. In both of these cases, these correspond to lower virial parameter estimates than that of the 3D calculation.

For all three clouds, the first line (at $t = 0$ Myr) shows a hook shape, which closely follows the line of $t = 0.1$ Myr, albeit with a slightly steeper hook transition. Also, the lines do not extend as far along the $x$-axis. However, what this clearly shows is a ``supported'' signature, from the outset, before the clouds have undergone any collapse. However, we know from the initial energy setup that there is no support in these clouds. We will discuss this further in \S\ref{subsec:why_increasing}. For time $t = 0.1$ Myr, we also see this line shows the characteristic hook upwards from $\log_{10} \Sigma_{\mathrm{thresh}}/\langle \Sigma \rangle \sim 0.25$ onwards. For the following two timesteps, $t~=~0.25$~Myr and $t~=~0.42$~Myr, for all clouds, we also see a hook shape, though it becomes less pronounced as the clouds evolve. By $t = 0.52$ Myr, corresponding to the initial free-fall times of the clouds, the DVA profile is much flatter, as is predicted for an already collapsed system.

In the final timestep, for cloud HDvir, we see this line trace $\alpha_{\mathrm{vir,obs}} \sim 1$, whereas for HDvir2, the DVA profile sits slightly lower, and is a bit noisier, but nonetheless much flatter than the initial profiles. For HDcold, at $t = 0.61$ Myr, we see the DVA profile gradually decrease out to $\log_{10} \Sigma_{\mathrm{thresh}}/\langle \Sigma \rangle \sim 1.3$, whereupon it begins to rise again. This decrease followed by growth in $\alpha_{\mathrm{vir,obs}}$ is also seen in the analytical solutions for collapsing polytropes, given in section 3.1.2 of \citet{Krumholz+25}. 

For the bottom panels of Figure~\ref{fig:VirialDiferencialAll}, we see a similar trend for $\langle \sigma / R_{\mathrm{eff}} \rangle$.
In both cases, for all clouds, the corresponding tracks bend upward relative to lines of constant virial parameter. This is precisely the qualitative behaviour that has been interpreted as the signature of a cloud whose dense internal structures are not undergoing global collapse, but in equilibrium. In the present case, however, the clouds are collapsing throughout their evolution. }}

We therefore see clearly that for our suite of clouds, as the evolution proceeds, the differential virial profiles become flatter over part of the sampled range in surface density, and the tracks in the virial diagram approach more closely the direction of constant $\alpha_{\rm vir,obs}$. } This behaviour resembles the temporal sequence found by \citet{Krumholz+25}, in which supported configurations produce upward-bending tracks and collapsing configurations produce flatter ones. The crucial difference is that, in our simulations, this sequence does not represent a transition from hydrostatic support to collapse. Instead, it occurs within a cloud that is gravity-dominated from the beginning. This result demonstrates that an increasing $\alpha_{\rm vir,obs}$ with $\Sigma$ is not a sufficient condition for excluding clouds undergoing collapse. The projected virial parameter can increase toward high surface density even when the dense gas belongs to a globally collapsing system. Conversely, the progressive flattening of the profile at later times does not necessarily mark the onset of collapse; it may instead reflect the continuing evolution of an already collapsing cloud. We will discuss the origin of this temporal sequence further in Section~\ref{Sec:Discussion}.

\section{Discussion}\label{Sec:Discussion}

The results presented in the previous section show that the differential virial analysis is not a unique indicator of the dynamical state of a molecular cloud. A cloud undergoing gravitational contraction from the outset can nevertheless exhibit an increasing $\alpha_{\rm vir,obs}$ toward high surface densities, followed by a progressive flattening of the differential virial profile as the collapse proceeds. This behaviour is qualitatively similar to that expected, in the interpretation of \citet{Krumholz+25}, for a sequence from supported to collapsing configurations. In our simulations, however, the underlying dynamical history is different: the cloud is not initially supported; it is gravity-dominated throughout its evolution.

As an extreme example, take the configuration at $t=0$, when the density is perfectly constant, and the cloud perfectly isothermal, which implies that there is no pressure gradient, and thus, there is no way in which the cloud can be in any type of equilibrium. Nonetheless, the thick solid lines in Fig.~\ref{fig:VirialDiferencialAll}, which correspond to $t=0$, exhibit the characteristic hook shape which have been interpreted as a `marker' of clouds that are not undergoing collapse because the pressure gradient is relevant \citep[see conclusions in]{Krumholz+25}. 

\subsection{Why can collapsing clouds show increasing virial parameters?}
\label{subsec:why_increasing}

The observational virial parameter in eq.~(\ref{eq:avir_obs}) is an observational estimator constructed from projected quantities that gives an insight on the boundedness of a three-dimensional structure. However, as we have shown, it does not unequivocally measure the actual instantaneous gravitational boundedness. In a collapsing cloud, the surface density, effective radius, and velocity dispersion all evolve simultaneously, and not necessarily in the same way at all column-density thresholds. Two of our clouds start from slightly subvirial turbulent velocity fields superposed on an initially nearly homogeneous density distribution, and are therefore not in hydrostatic equilibrium. The case for HDcold is even more extreme. Even before strong local collapse develops, the projected column density increases toward the centre because lines of sight through the inner regions sample a larger fraction of the cloud mass. The velocity fields, however, were initialized over the full cloud volume, and therefore contain modes whose amplitudes are set by the gravitational scale of the whole system, not by the mass enclosed within each smaller projected aperture. Consequently, when the same large-scale turbulent motions are compared with the smaller masses and radii associated with high-column-density regions, the inferred $\alpha_{\rm vir,obs}$ can increase toward high $\Sigma$. We can see this when analysing the relationship of $\sigma$, $R_{\mathrm{eff}}$, and $\Sigma$ with respect to $\alpha_{\mathrm{vir, obs}}$. From Equation~\ref{eq:avir_obs}, we know $\alpha_{\mathrm{vir,obs}} \propto \sigma^2/(R_{\mathrm{eff}}\Sigma)$. In our case, $\sigma$ is initially set across the whole cloud. Therefore, the change in $\alpha_{\mathrm{vir,obs}}$ with radius is governed by the effective radius we are probing, and the internal density distribution of the cloud. For a flat density profile, which corresponds to the early stages of our clouds' evolution, $\Sigma$ rises gradually, whilst $R_{\mathrm{eff}}$ shrinks much more rapidly. As such, the change in $\alpha_{\mathrm{vir}}$ comes purely from the system geometry. If the density profile was closer to the singular isothermal sphere ($\rho\propto r^{-2}$), $\Sigma$ would also rise sharply, keeping $\alpha_{\mathrm{vir,obs}}$ flat.} 

Therefore, for the early phases of the simulation, DVA produces an apparently supervirial differential profile even though the clouds as a whole are gravitationally dominated and undergoing collapse. Thus, the upward trend of $\alpha_{\rm vir,obs}$ at early times does not indicate that the dense gas is supported against gravity; it reflects a difference between the spatial scale of the velocity field and the smaller projected scales on which the virial estimator is evaluated.

The same result can be shown in 3D. In this case, the virial parameter is given by
\begin{equation}
    \alpha_{\mathrm{vir}} = a\frac{\sigma^2 R}{GM(R)}
    \label{eq:avir_3d}
\end{equation}
where $a$\footnote{For a homogeneous sphere, $a=5/3$}  is a prefactor set by the geometry of the system, and $M(R)$ is the mass enclosed at a radius $R$. For a spherical cloud, $M(R) \propto R^3\rho(R)$, and so the dependency of $\alpha_{\mathrm{vir}}$ on $R$ is set by the relationship between density and radius, $\rho \propto R^{-k}$. For a flat density profile, $k = 0$, and so $\alpha_{\mathrm{vir}} \propto \sigma^2 R^{-2}$. For a constant $\sigma$, the virial parameter therefore rises as $R^{-2}$. However, for a centrally condensed density configuration, but constant $\sigma$, the dependency on $R$ can be erased, and return a much flatter relationship between $\alpha_{\mathrm{vir}}$ and $R$. For a singular density profile, $\rho \propto R^{-2}$, leading to no dependence on radius for the virial parameter. Once again, for a constant $\sigma$, this leads to a constant $\alpha_{\mathrm{vir}}$ throughout the cloud. Crucially, however, the actual ratio between the kinetic and potential energies of the cloud has not changed throughout this example, and in both cases, the cloud will collapse. This is because the argument centres on the stability of the cloud against collapse on a global scale, not on an internal one.

\begin{figure}
    \centering
    \includegraphics[width=\linewidth]{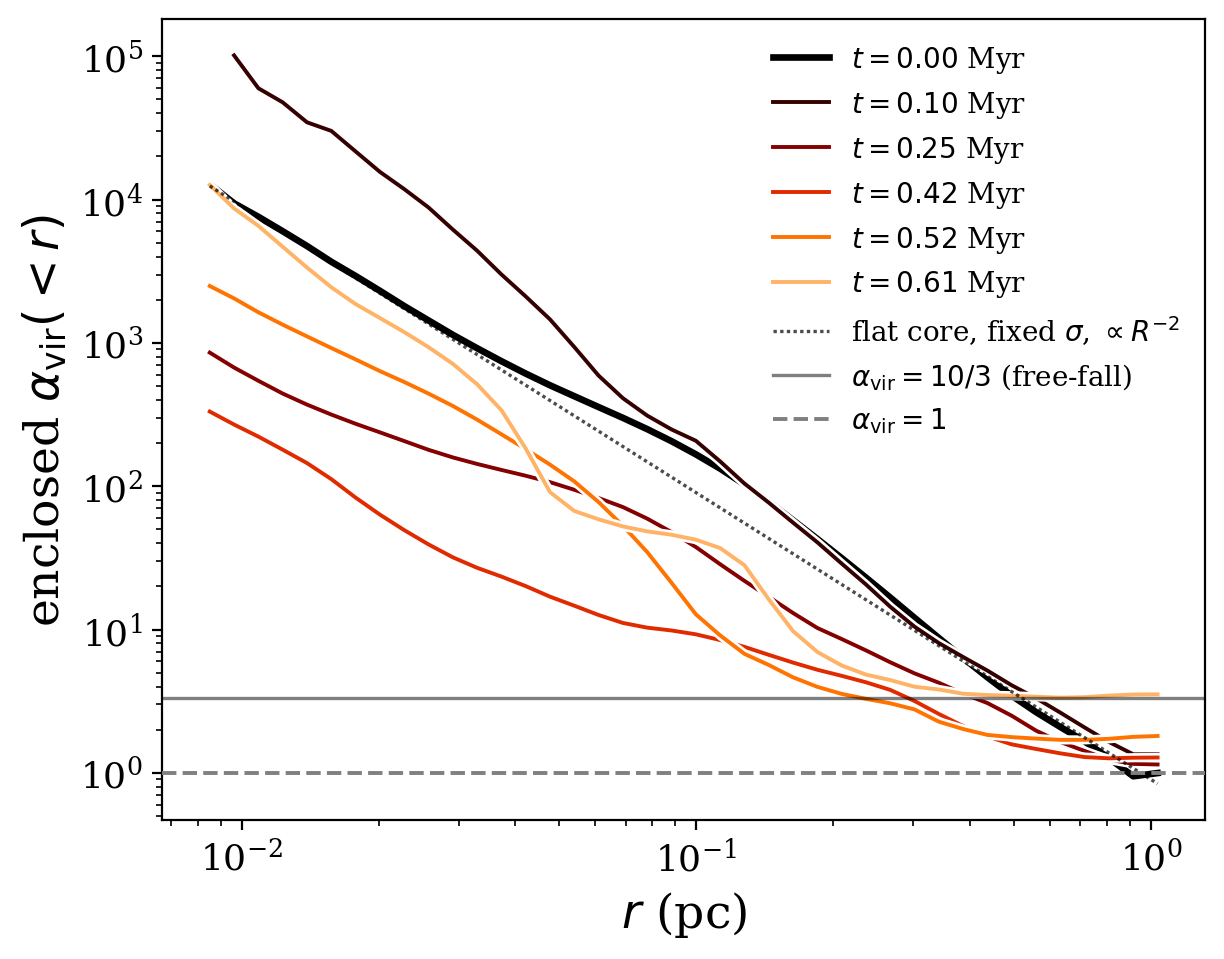}
    \caption{Enclosed virial parameter as a function of radius within the cloud, shown for HDvir2, shown for different cloud evolutionary states. Lighter lines are earlier times and darker later times. The thick black line shows the relationship between $\alpha_{\mathrm{vir, enc}}$ and $r$ at the onset of the simulations. Here, the line traces closely the predictions of a flat core, with a constant velocity dispersion throughout. As the cloud evolves, the lines depart from this relationship as density and velocity structures form. In the final timestep, we see the outer regions of the cloud have a flatter $\alpha_{\mathrm{vir, enc}}$, consistent with the predictions of free-fall.}
    \label{fig:sshell}
\end{figure}

As an example, for cloud HDvir2, we have plotted the enclosed virial parameter at multiple shells with radii $r_{\mathrm{shell}} \leq R_{\mathrm{cloud}}$, for each different time evaluated in our DVA analysis in Figure~\ref{fig:sshell}, as well as at $t~=~0$~Myr, before the cloud has evolved at all. Here, especially at $t=0$, we see the effect of a broadly consistent $\sigma$, which was set on the spatial scale of the cloud and independently of radius. As such, $\alpha_{\mathrm{vir}}$ rises as $R^{-2}$. However, as we have shown repeatedly throughout, this cloud was never stable against collapse. 

As the cloud evolves, the profiles depart from this baseline as central structure develops and coherent motions grow. We even see that by $t = 0.61$ Myr, material within $0.4 \leq r \leq 1.0$ pc has flattened to an $\alpha_{\mathrm{vir}} \sim 10/3$, which is characteristic of free-fall.\footnote{From infinity, the free-fall velocity is $v_{\mathrm{ff}} = \sqrt{2GM/R}$, which, when plugged into Equation~\ref{eq:avir_3d} gives $\alpha_{\mathrm{vir}} = 10/3$. } As such, we propose that what DVA actually informs you about is the internal structure of a system, within a set velocity scale. We discuss this more clearly in \S\ref{subsec:DVAuse}.

It is worth discussing the initial tests done for the polytropes in  \citet[][see their Section 3.1]{Krumholz+25}. These authors perform the DVA of three profiles: a critical non-singular core, a supercritical non-singular core, and a singular core. Both non-singular cases therefore have a central region with a flat density profile. Based on this setup, however, the supercritical core, whilst instantaneously supported, is in fact unstable to collapse if left to evolve. Yet, from Fig.~1 in \citet{Krumholz+25} , this core still returns the characteristic ``hook'' ascribed to support, however, the position of the hook has shifted rightwards along the $x$-axis, to a higher column density. The difference between the two cases is the radius out to which the authors integrate. For the critical core, the integration is done out to the point at which the core becomes unstable, whereas for the supercritical core, the integral goes out to where the centre-to-edge density ratio reaches 100. For two spheres of identical radius, $R$, the supercritical core has moved `towards' a singular profile. Therefore, the shape of the $\alpha_{\mathrm{obs,vir}} - \Sigma$ relation describes how close an object is to singularity, rather than whether it is supported or not. This is in line with the discussion above, following the relationship between internal density profiles of the clouds, and their transition from flat to centrally condensed.

So far, we have shown that for a projected region selected above a surface-density threshold, an increase of $\alpha_{\rm vir,obs}$ with $\Sigma$ means only that the projected kinetic term $\sigma^2/R_{\rm eff}$ increases faster than $\Sigma$. It does not specify the physical origin of the velocity dispersion. However, we have done this all in the context of our isolated clouds, which provide a simplistic set of initial conditions to test this method on. The situation for real molecular clouds is yet more complex. For these systems, the measured linewidths may include random turbulent motions, as well as infall, accretion flows, shear, unresolved substructure, and projection of coherent velocity gradients. Several of these contributions are expected on clumps within a collapsing cloud. All of this makes untangling the exact source of DVA profiles much harder.

It is worth noticing that in a hierarchically collapsing cloud, different regions and scales may become unstable at different times. Indeed, large-scale contraction may begin before the densest projected substructures have reached their own local collapse phase \citep{Naranjo-Romero+15, Vazquez-Semadeni+19}. During this stage, dense regions can be embedded in a globally collapsing flow while still exhibiting large internal velocity dispersions or complex projected morphologies. Their observational virial parameters may therefore appear elevated even though the cloud as a whole is not supported against collapse. This point is important because the differential virial parameter has been used as if it were a local collapse indicator. It is not. A high value of $\alpha_{\rm vir,obs}$ in a column-density-selected structure does not necessarily mean that the gas is locally supported against gravity. It may instead indicate that the selected projected region contains a mixture of dynamically distinct components, coherent motions, or unresolved velocity gradients generated by the collapse itself.

A central ambiguity in virial analyses is the distinction between virial-like energy balance and hydrostatic equilibrium. A system in hydrostatic equilibrium may have kinetic, thermal, magnetic, and surface terms that balance self-gravity. However, a collapsing system also develops a kinetic energy comparable to its gravitational energy because gravitational contraction converts potential energy into motion, as shown by \citet{Vazquez-Semadeni+07} for a gas system, and by \citet{Noriega-Mendoza_Aguilar18} for a stellar system. In that case, the appearance of virial balance is a consequence of the collapse, not evidence against it.

This distinction has been emphasized in previous numerical and analytical work on gravity-driven cloud evolution \citep{Vazquez-Semadeni+07, Ballesteros-Paredes+11a, Ballesteros-Paredes+18, Noriega-Mendoza_Aguilar18}. In such models, molecular clouds do not need to be long-lived, turbulence-supported objects in order to display virial parameters of order unity. Their linewidths can arise naturally from accretion, contraction, and hierarchical collapse.

The same caveat applies to DVA. A flat differential virial profile may be compatible with global collapse, but a non-flat profile does not automatically imply support. The virial diagram does not encode the sign of the motions, nor does it separate pressure-like random motions from coherent converging flows. Therefore, a DVA profile alone is not enough to predict the stability of a system on a global scale.

\subsection{What does DVA inform us about?}
\label{subsec:DVAuse}
Although our analysis indicates that the DVA is not a good discriminator or `marker' for whether the cloud intrinsically has a pressure gradient opposing gravitational collapse, it instead suggests that DVA measures \textit{how centrally concentrated a system is (i.e., how close the system is to singular). For a globally collapsing cloud, this maps directly onto the evolutionary stage, with a strong hook indicating a structurally young, near-uniform density distribution, whereas a flat DVA profile indicates a more structurally advanced, condensed system. Therefore, when utilised as a probe of density structure, the DVA technique can be genuinely informative about the degree to which a system has evolved (when combined with constraints on the velocity field within the analysed region). We suggest future work, analysing simulations with various different initial conditions, configurations, and scales would be useful to validate this hypothesis, but this is currently beyond the scope of this paper. }

\section{Conclusions}\label{Sec:Conclusions}
{
The dynamical state of molecular clouds is a key element of any star formation theory. DVA suggests that the dynamical state of a system can be determined by analysing the evolution of the observed virial parameter as a function of column density -- thus showing the internal evolution of the virial parameter within the object. The key assumption here is that each dynamical state must have its own signature form; in this case, a globally collapsing system will produce a constant virial parameter as you probe the inner regions, whereas a hydrostatic system will show an increasing virial parameter as you move inwards.}

{Within this framework, we have performed a DVA on a set of simulations of collapsing clouds of mass $1000~M_\odot$ and radius 1~pc, over the course of $\sim 1.2 t_{\mathrm{ff}}$. Two of these clouds begin initially slightly sub-virial, whilst the third begins very subvirial. As such, the clouds are always in a state of collapse. In contrast to \citet{Krumholz+25}, we see the same initial `hook' signature during the beginning phases of collapse, followed by a transition to an approximately constant value of $\alpha_{\mathrm{vir,obs}}$ over the course of one free-fall time. From this, we conclude that:}

{
\begin{enumerate}
 \item the increasing virial parameter with column density, is not unique to a hydrostatic system.
 
  \item The transition from increasing to flat virial parameter with column density 
  is not unique to a supported system transitioning to collapse.

  \item {This is because $\alpha_\mathrm{vir, obs}$ depends on three parameters, $R_{\mathrm{eff}}$, $\Sigma$, and $\sigma_{\mathrm{LOS}}$, which evolve independently during the collapse of a cloud. As such, different combinations of these three parameters may produce the same shape, whilst representing different physical states for the cloud (e.g., different contributions to the LOS-velocity, from convergent flows to turbulent internal support). We demonstrate this clearly by repeating the same DVA on our own collapsing systems.}

  \item For a set velocity scale, DVA tells you about the internal density profile of a system, but cannot be used as a predictor of overall stability.
  
  \item {As such, making a definitive conclusion about the dynamical state of molecular clouds cannot be done through this analysis alone. To do so would require a more thorough analysis of the quantities which determine $\alpha_{\mathrm{vir, obs}}$. However, we have shown that a reinterpretation of the results of DVA can provide an informative diagnostic for the evolutionary state of a cloud, over a set velocity scale.}
\end{enumerate}}

\section*{Acknowledgements}

KK acknowledges postdoctoral fellowship from UNAM-DGAPA.
JBP acknowledges financial support from the UNAM-DGAPA-PAPIIT IN110026 grant.
This work made use of dedicated computing resources developed and maintained by the IRyA's Computing Department. The authors gratefully acknowledge the contributions of Daniel Díaz-González, Miguel Espejel, Alfonso Ginori González, and Gilberto Zavala.
In addition, we acknowledge the support of the IRyA Library. In particular, we thank Leonardo Arroyo for providing bibliographic resources that contributed to this work.
Finally, the authors gratefully acknowledge the support of the IRyA Outreach personnel. In particular, we thank René Ortega for his assistance in the communication and dissemination of this work.

\section*{Data Availability}
 
The data underlying this article will be shared on reasonable request to the corresponding author.



\bibliographystyle{mnras}
\bibliography{MisReferencias} 





\bsp	
\label{lastpage}
\end{document}